\documentclass[12pt]{article}

\usepackage[paper=a4paper, top=2cm, bottom=3cm, left=2.5cm, right=2.5cm]{geometry}

\usepackage{ucs}
\usepackage[utf8x]{inputenc}
\usepackage[english]{babel}
\usepackage{underscore}	  
\usepackage{subfigure}
\usepackage{verbatim}
\usepackage{float}
\usepackage{booktabs}     
\usepackage{parskip}
\usepackage{caption}
\usepackage{hyperref}
\usepackage{amsmath}
\usepackage{mathabx}

\usepackage{graphicx}
\DeclareGraphicsExtensions{.pdf,.png,.jpg}

\usepackage{color}
\definecolor{mygray}{rgb}{0.5,0.5,0.5}

\usepackage{listings}
\newcommand{\file}[1]{\texttt{\small #1}}

\title{SandBlaster: Reversing the Apple Sandbox}

\author{\small{Răzvan Deaconescu, Luke Deshotels, Mihai Bucicoiu,}\\
  \small{William Enck, Lucas Davi, Ahmad-Reza Sadeghi}\\
\footnotesize \emph {razvan.deaconescu@cs.pub.ro, ladeshot@ncsu.edu, mihai.bucicoiu@cs.pub.ro,}\\
\footnotesize \emph {whenck@ncsu.edu, lucas.davi@trust.cased.de, ahmad.sadeghi@trust.cased.de} }

\date{\today}

\begin{document}

\maketitle

\begin{abstract}
In order to limit the damage of malware on Mac OS X and iOS, Apple uses sandboxing, a kernel-level security layer that provides tight constraints for system calls. Particularly used for Apple~iOS, sandboxing prevents apps from executing potentially dangerous actions, by defining rules in a sandbox profile. Investigating Apple's built-in sandbox profiles is difficult as they are compiled and stored in binary format. We present SandBlaster, a software bundle that is able to reverse/decompile Apple binary sandbox profiles to their original human readable SBPL (\textit{SandBox Profile Language}) format. We use SandBlaster to reverse all built-in Apple iOS binary sandbox profiles for iOS 7, 8 and 9. Our tool is, to the best of our knowledge, the first to provide a full reversing of the Apple sandbox, shedding light into the inner workings of Apple sandbox profiles and providing essential support for security researchers and professionals interested in Apple security mechanisms.

\end{abstract}

\section{Introduction}
\label{sec:introduction}
Apple iOS is the second most popular operating system for mobile devices~\cite{ios-popularity}. It provides a variety of security mechanisms~\cite{iOS9White} targeted at providing privacy for its users. One such mechanism is the Apple sandbox, implemented as part of the XNU kernel used both by iOS and Mac OS X. The Apple sandbox employs the TrustedBSD Mandatory Access Control framework and confines an app to only a subset of actions defined as part of a \textit{sandbox profile}. The sandbox is used on iOS and Mac OS X to reduce the damage of malware or an exploited app or system process.

The information provided by Apple regarding the sandbox~\cite{apple-app-sandbox-design-guide} is insufficient for security researchers and professionals to gain an in-depth overview. Moreover, the built-in sandbox profiles used by iOS are stored in binary format. Sandbox profiles are initially written as a human readable set of rules using a Scheme-like language (dubbed SBPL, \textit{SandBox Profile Language}). Built-in iOS sandbox profiles are then compiled to a binary format and stored in the iOS filesystem. Binary sandbox profiles consist of serialized graphs, thus making queries on them more efficient than on human readable profiles. As the binary format is prohibitively difficult to analyze, one needs to reverse the built-in Apple iOS sandbox profiles.

We present SandBlaster, a software bundle that reverses binary sandbox profiles into their original human readable SBPL (\textit{SandBox Profile Language}) format. Part of our work is based on previous work by Dionysus Blazakis~\cite{Blazakis11theapple} and Stefan Esser~\cite{esserSandbox}. Dionysus Blazakis wrote the original reverser for iOS~4~\footnote{\url{https://github.com/dionthegod/XNUSandbox}}. With minor updates the reverser can be used on iOS~5 and iOS~6. Blazakis' earlier work featured the reversal of binary sandbox profiles to an intermediary human readable format, different to SBPL. Starting with iOS~7, Apple updated the binary sandbox format further complicating the reversing process. Stefan Esser used Blazakis' ideas and updated the implementation for iOS~7, 8 and 9~\footnote{\url{https://github.com/sektioneins/sandbox_toolkit}}. The output from Esser's tools is a graph representation of sandbox profiles. The step of converting this graph representation to SBPL format is absent and so is the possibility of modifying, testing and evaluating built-in sandbox profiles.

SandBlaster consists of original software, existing tools and scripts that automate the process of reversing binary sandbox profiles to SBPL format. By incorporating existing reverse engineering tools, we are able to provide the initial SBPL format for built-in iOS binary sandbox profiles. We use firmware files (\texttt{.ipsw}) published by Apple with no need for a device running iOS. We reversed all built-in binary sandbox profiles in all major versions for iOS~7, iOS~8 and iOS~9. Reversed SBPL format profiles, particularly the default sandbox profile for third party apps (dubbed \textit{container}), allow critical analysis on the security and privacy features provided by the Apple sandbox for iOS apps.


\section{Apple Sandbox Profiles}
\label{sec:sandbox-profiles}
As previously mentioned, the Apple sandbox implementation is part of the XNU kernel, the kernel used by both iOS and Mac OS X. The sandbox confines a given process to a set of actions (i.e. system calls and system call arguments) defined in a sandbox profile. As they use the same kernel, XNU, the iOS and Mac OS X sandbox implementations are similar; investigating the Mac OS X implementation provides insight into the iOS implementation as well, and vice versa. For SandBlaster our focus is on iOS as it makes heavy use of sandboxing with every third party application and many system applications making use of sandbox profiles.

\subsection{Overview of an Apple Sandbox Profile}
\label{subsec:sandbox-profile-overview}

An Apple sandbox profile is a list of rules used by the XNU kernel to make access control decisions for actions invoked by third party apps or system processes. Listing~\ref{lst:sandbox-profile-overview} shows a high level view of the sandbox profile format.

\begin{lstlisting}[caption={High level view of sandbox profile},label={lst:sandbox-profile-overview}]
sandbox_profile
    (* $\drsh$ *)operation, decision
        (* $\drsh$ *)filter
        (* $\drsh$ *)filter
        (* $\drsh$ *)filter
        (* $\dots$ *)
    (* $\drsh$ *)operation, decision
        (* $\drsh$ *)filter
        (* $\drsh$ *)filter
        (* $\drsh$ *)filter
        (* $\dots$ *)
    (* $\drsh$ *)operation, decision
        (* $\drsh$ *)filter
        (* $\drsh$ *)filter
        (* $\drsh$ *)filter
        (* $\dots$ *)
    (* $\dots$ *)
\end{lstlisting}

As shown above, there are three key components for a sandbox profile: operations, decisions and filters.

An \textbf{operation} (e.g., \texttt{file-read}, \texttt{mach-lookup}, \texttt{network-outbound}) is an abstraction of a system call or functionality invoked by a process and checked by the kernel. The kernel decides whether the process making the call is allowed or denied the operation by checking its \textbf{decision} and its \textbf{filter}. An operation is permitted if its filter is matched and its decision is \textit{allow}; otherwise the operation is denied.

Simple rules inside the sandbox profile state that certain operations are always allowed or denied irrespective of other information (such as the file name, the process ID, or the remote socket address). Most often, though, the actual decision for allowing or denying an operation rests on verifying that other criteria are met. These criteria are implemented through \textbf{filters}. A filter is defined in the sandbox profile as a key-value pair (e.g., \texttt{literal "/bin/secret.txt"}, \texttt{vnode-type REGULAR-FILE}, \texttt{remote tcp "localhost:22"}). Keys and values in the pair are strings, numbers and other constructs that are translated into a number inside the serialized filter; some example mappings are shown in Table~\ref{tab:sample-filter-mappings}. The decision is applied only if the given filters are matched.

\begin{table}[h]
  \begin{center}
  \begin{tabular}{lll}
    \toprule
    \textbf{Binary Filter Key \& Value} & \textbf{SBPL Equivalent} \\
    \midrule
    \texttt{0x01} \& \texttt{0x0002} & \textit{literal "/path/to/file"} \\
    \texttt{0x1d} \& \texttt{0x0001} & \textit{vnode-type REGULAR-FILE} \\
    \texttt{0x0e} \& \texttt{0x0004} & \textit{target children} \\
    \texttt{0x0c} \& \texttt{0x0001} & \textit{socket-type SOCK_STREAM} \\
    \bottomrule
  \end{tabular}
  \end{center}
  \caption{Sample Filter Mappings}
  \label{tab:sample-filter-mappings}
\end{table}

Sandbox profiles are defined in the SandBox Profile Language (SBPL) format. SBPL is similar to a functional programming language format and defines the operations, decisions and filters. Listing~\ref{lst:sandbox-snippets} shows two possible lines in a sandbox profile. The first line denies the \texttt{file-read*} operation only if the path is \texttt{/bin/secret}. The second line allows the \texttt{network-outbound} operation only if the remote TCP endpoint is running on \texttt{localhost} on port \texttt{22}.

\begin{lstlisting}[caption={Sample lines in a sandbox profile},label={lst:sandbox-snippets}]
(deny file-read* (literal "/bin/secret"))
(allow network-outbound (remote tcp "localhost:22"))
\end{lstlisting}

Each sandbox profile has a \texttt{default} operation. The decision for the \texttt{default} operation is taken in case there is no match for other operations; e.g. if no filters of the \texttt{file-read*} operation are matched, then the decision for the \texttt{default} operation is taken. For almost all iOS sandbox profiles the \texttt{default} operation uses the \textit{deny} decision~\footnote{In iOS 9.3.1, 103 out of the 121 built-in sandbox profiles use the \textit{deny} decision for the \texttt{default} operation}, making the operation rules a whitelist. In other words, any operation that is not explicitly allowed by the profile will be denied.

Listing~\ref{lst:simplified-sandbox-profile} is a simplified example of a sandbox profile in SBPL format using the \texttt{default} and \texttt{file-read*} operations and two filters: \texttt{literal} and \texttt{regex}. The rules inside the profile allow the sandboxed application to read any file in the \file{/bin/} directory except the \file{/bin/secret.txt} file. Whenever a file is read, the sandbox engine matches the filters for the \texttt{file-read*} operation:
\begin{enumerate}
  \item If the file name is \file{/bin/secret.txt}, the operation is denied, due to the rule in line 2.
  \item If the file name matches the regular expression \texttt{/bin/*} (i.e. the file is part of the \file{/bin/} directory), the operation is allowed, due to the rule in line 3.
  \item In any other situation, the operation is denied.
  \item Any operations other than \texttt{file-read*} are by default denied, due the rule in line 1.
\end{enumerate}

\begin{lstlisting}[caption={Sample lines in a sandbox profile},label={lst:simplified-sandbox-profile}]
(deny default)
(deny file-read* (literal "/bin/secret.txt"))
(allow file-read* (regex #"/bin/*"))
\end{lstlisting}

fG!'s unofficial documentation~\cite{fgSandboxGuide} provides a reference of common operations and filters and an anatomy of sandbox profiles. The reference is contemporary to Blazakis' earlier work; it investigates Mac OS X 10.6.8 Snow Leopard sandboxing implementation, with 59 operations covered. However, new operations and filters were introduced in each iOS version with an increase from 88 operations in iOS~6, to 114 operations in iOS~7 and iOS~8, to 119 operations in iOS~9.0 and 125 operations in iOS~9.3.

\subsection{Metafilters}
\label{subsec:metafilters}

Metafilters apply logical operations (e.g., and, or, not) to filters. We call them metafilters because they aggregate other (meta)filters and because they are not defined in the usual key-value format.

The \texttt{require-any} metafilter is the equivalent of a \textbf{logical or} between the different filters. In Listing~\ref{lst:require-any} the \texttt{file-read*} operation is allowed if \textbf{any} of the two filters is matched: either the \textit{regex} filter on line 3 or the \textit{vnode-type} filter on line 4.

\begin{lstlisting}[caption={Using the require-any metafilter},label={lst:require-any}]
(allow file-read*
    (require-any
        (regex #"/bin/*")
        (vnode-type REGULAR-FILE)))
\end{lstlisting}

To match all the filters in a list the \texttt{require-all} metafilter is applied. The \texttt{require-all} metafilter is the equivalent of a \textbf{logical and} between the different filters. In the case of Listing~\ref{lst:require-all} the operation is allowed if both the \texttt{regex} and the \texttt{vnode-type} filters are matched.

\begin{lstlisting}[caption={Using the require-all metafilter},label={lst:require-all}]
(allow file-read*
    (require-all
        (regex #"/bin/*")
        (vnode-type REGULAR-FILE)))
\end{lstlisting}

The \texttt{require-not} metafilter matches the negation of a filter. In the case of Listing~\ref{lst:require-not} the operation is allowed only if the \texttt{vnode-type} filter is \textbf{not} matched.

\begin{lstlisting}[caption={Using the require-not metafilter},label={lst:require-not}]
(allow file-read*
    (require-not
        (vnode-type REGULAR-FILE)))
\end{lstlisting}

An operation may use multiple filters, as is the case with the operation in Listing~\ref{lst:multiple-filters}. In this case the \texttt{file-read*} operation is allowed if either the \texttt{regex} or the \texttt{vnode-type} filter is matched. This is equivalent to using the \texttt{require-any} metafilter, as shown in Listing~\ref{lst:require-any}.

\begin{lstlisting}[caption={Using multiple filters},label={lst:multiple-filters}]
(allow file-read*
    (regex #"/bin/*")
    (vnode-type REGULAR-FILE))
\end{lstlisting}

Metafilters may be nested resulting in complex rules inside the sandbox profiles.

Metafilters are not handled by Esser's tools, but are handled by SandBlaster. SandBlaster does a full reversing of a binary sandbox profile to the initial SBPL format including the \textit{require-not}, \textit{require-any} and \textit{require-all} metafilters. We detail this in Subsection~\ref{subsubsec:reverse-require-not} and Subsection~\ref{subsubsec:reverse-nested-rules}.

\subsection{Storing Sandbox Profiles in iOS}
\label{subsec:storing-sandbox-profiles}

Information on storing binary sandbox profiles in iOS and the binary format is not provided by Apple. We extracted information below through reverse engineering, taking into account previous work by Blazakis and Esser. We then integrated this information into SandBlaster, as we show in Section~\ref{sec:methodology} and Section~\ref{sec:internals}.

For efficient queries, sandbox profiles are stored as binary blobs in iOS. Between iOS~5 and iOS~8, binary blobs for sandbox profiles were stored in the \texttt{/usr/libexec/sandboxd} file; between iOS~2 and iOS~4 and starting again with iOS~9, sandbox profiles are stored in the sandbox kernel extension (\texttt{com.apple.security.sandbox}). Thus, reversing the sandbox binary blobs to their original SBPL format requires getting access to the \texttt{/usr/libexec/sandboxd} file or the sandbox kernel extension, extracting the binary blobs and then reversing the binary blobs.

\begin{table}[h]
  \begin{center}
  \begin{tabular}{lll}
    \toprule
    \textbf{iOS Version(s)} & \textbf{Storage File} & \textbf{Storage Type} \\
    \midrule
    iOS~2 \ldots iOS~4 & \texttt{com.apple.security.sandbox} & separated \\
    iOS~5 \ldots iOS~8 & \texttt{/usr/libexec/sandboxd} & separated \\
    iOS~9 & \texttt{com.apple.security.sandbox} & bundled \\
    \bottomrule
  \end{tabular}
  \end{center}
  \caption{Storage Location and Storage Type of Binary Sandbox Profiles}
  \label{tab:sandbox-profile-store}
\end{table}

Until iOS~8, each sandbox profile was stored in an isolated binary blob; reversing a sandbox profile requires extracting its binary blob and reversing it. Starting with iOS~9, all sandbox profiles are bundled together in a single binary blob; reversing any sandbox profile means working with the binary blob bundle. In Table~\ref{tab:sandbox-profile-store} we show the storage location and the storage type for sandbox profiles depending on the iOS version.

\begin{figure}[h]
  \begin{center}
    \includegraphics[width=0.6\columnwidth]{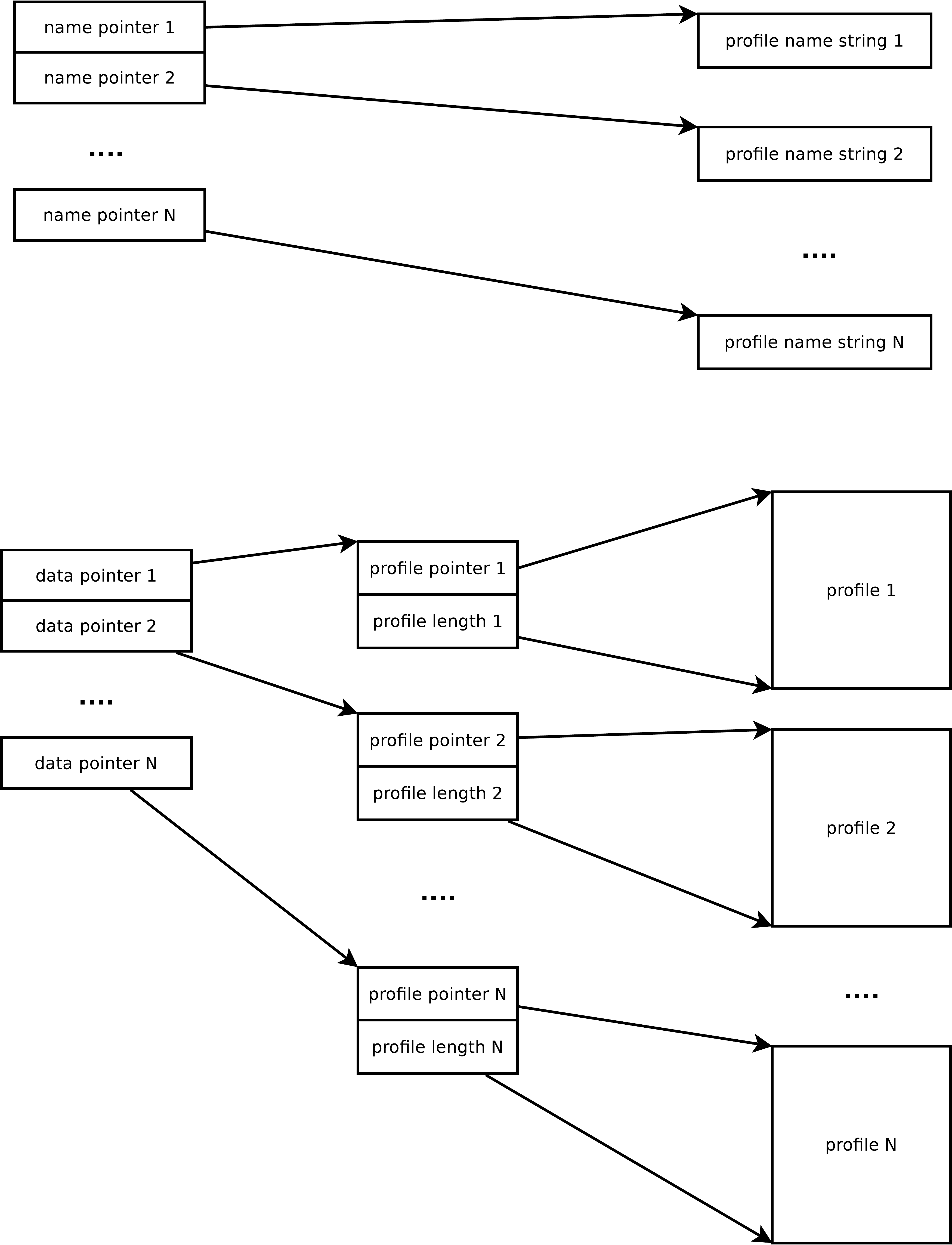}
  \end{center}
  \caption{Storing Sandbox Profile Information in iOS $<$ 9}
  \label{fig:sandbox-profile-store}
\end{figure}

Figure~\ref{fig:sandbox-profile-store} shows how sandbox profile information is stored in a separated storage in iOS $<$ 9. Stefan Esser's extraction tool considers this storage format and extracts sandbox profiles. In the figure, \texttt{name pointer 1} points to the string representing the name of the first profile. The contents of the first profile (binary blob) is pointed to by the \texttt{profile pointer 1} with the length determined by \texttt{profile length 1}, both of which are pointed to by \texttt{data pointer 1}. The same goes for the name and contents of the second profile and so on.

For iOS~9 all sandbox profile information (pointers, names and profile data) are bundled together. Given a \texttt{com.apple.security.sandbox} extension file we look for the bundle header and copy the extension file contents from that point until the end. We show more on the format of the bundled sandbox profiles in the next section.

SandBlaster reverses sandbox profiles in both formats (separated or bundled) as shown in Section~\ref{sec:internals}.

\subsection{The Binary Format of Sandbox Profiles}
\label{subsec:binary-format}

\begin{figure}[h]
  \begin{center}
    \includegraphics[width=0.5\columnwidth]{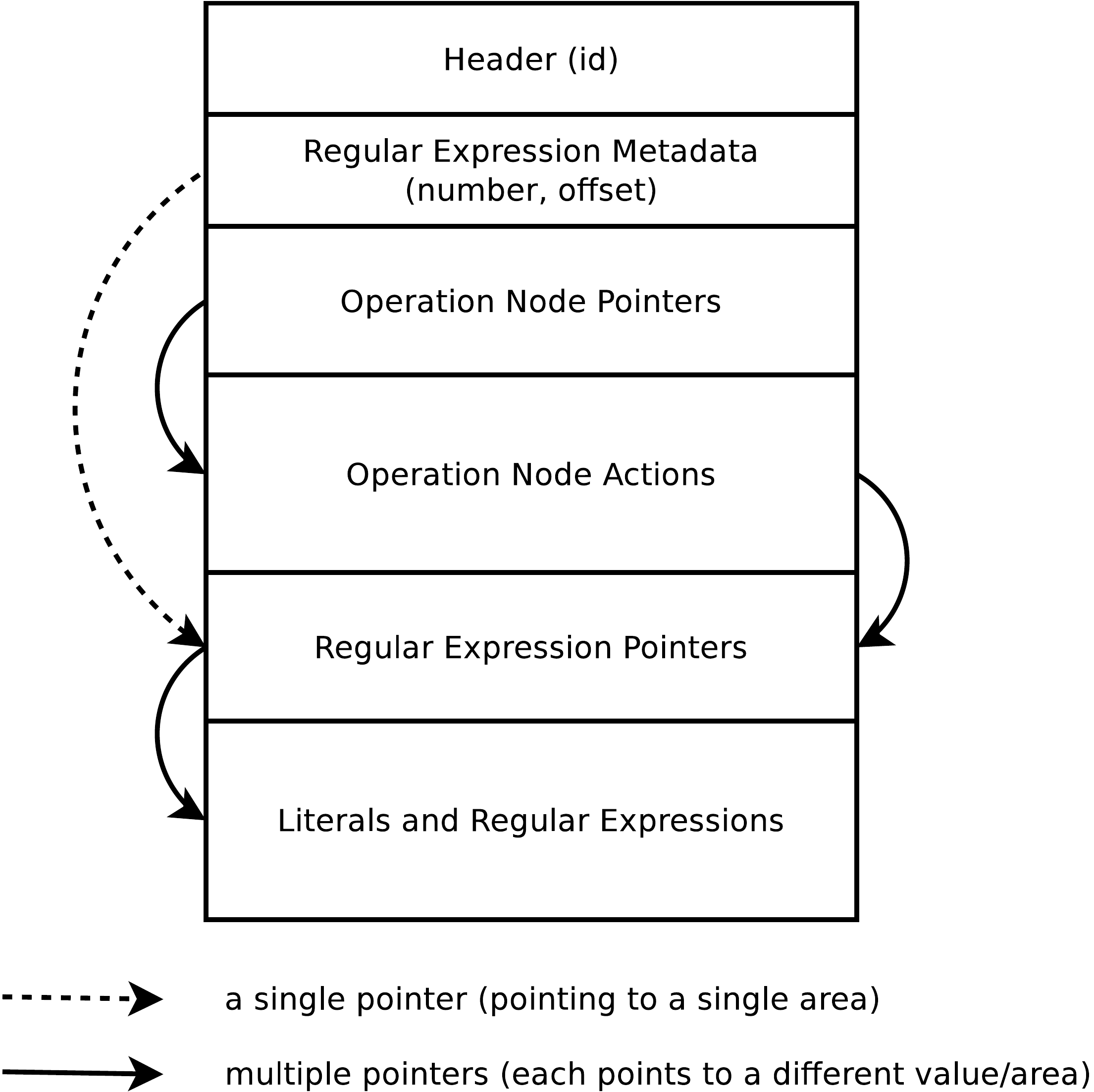}
  \end{center}
  \caption{Overview of the Binary Format of a Sandbox Profile (iOS $<$ 9)}
  \label{fig:binary-format}
\end{figure}

Figure~\ref{fig:binary-format} shows the sections inside a binary Apple sandbox profile file for iOS~7 and iOS~8~\footnote{We didn't investigate the format for iOS $<$ 7 but we expect it to be similar, if not identical.} in which each sandbox profile is stored separated from the others. Arrows in the figure represent pointers that reference data in other sections~\footnote{The section names in Figure~\ref{fig:binary-format} are defined by us, they are not standard (i.e. they are not documented by Apple).}. For example, there are pointers in the Operation Node Pointers section that reference values in the Operation Node Actions section.

The \textit{Operation Node Pointers} section is an array of offsets (pointers) to entries in the \textit{Operations Node Actions} sections. There is a pointer for each \textit{operation}; such that the array size (and the space occupied by the \textit{Operation Node Pointers} section) is equal to the number of available operations for the particular iOS version.

The most important part in the binary format is the \textit{Operation Node Actions} section. This is the serialized form of the \textit{filters} in the original SBPL format. An entry in the \textit{Operation Node Actions} section corresponds to a filter and often a decision (\textit{deny} or \textit{allow}). If it's not a decision, then it's a link to another serialized filter to be processed. We discuss more on this in Section~\ref{sec:internals}.

Filters may use regular expression patterns or literals. This is common for operations that deal with the file system. Filters that use regular expressions or literals use an index in an array stored in the \textit{Regular Expression Pointers} section; each item in the array is an offset to the actual literal or regular expression. Regular expressions are stored in a serialized form (similar to filters) in another section at the end of the binary profile (which we named \textit{Literals and Regular Expressions}), as shown in Figure~\ref{fig:binary-format}. \textit{Literals} are strings that are stored in the same section with regular expressions.

\begin{figure}[h]
  \begin{center}
    \includegraphics[width=0.6\columnwidth]{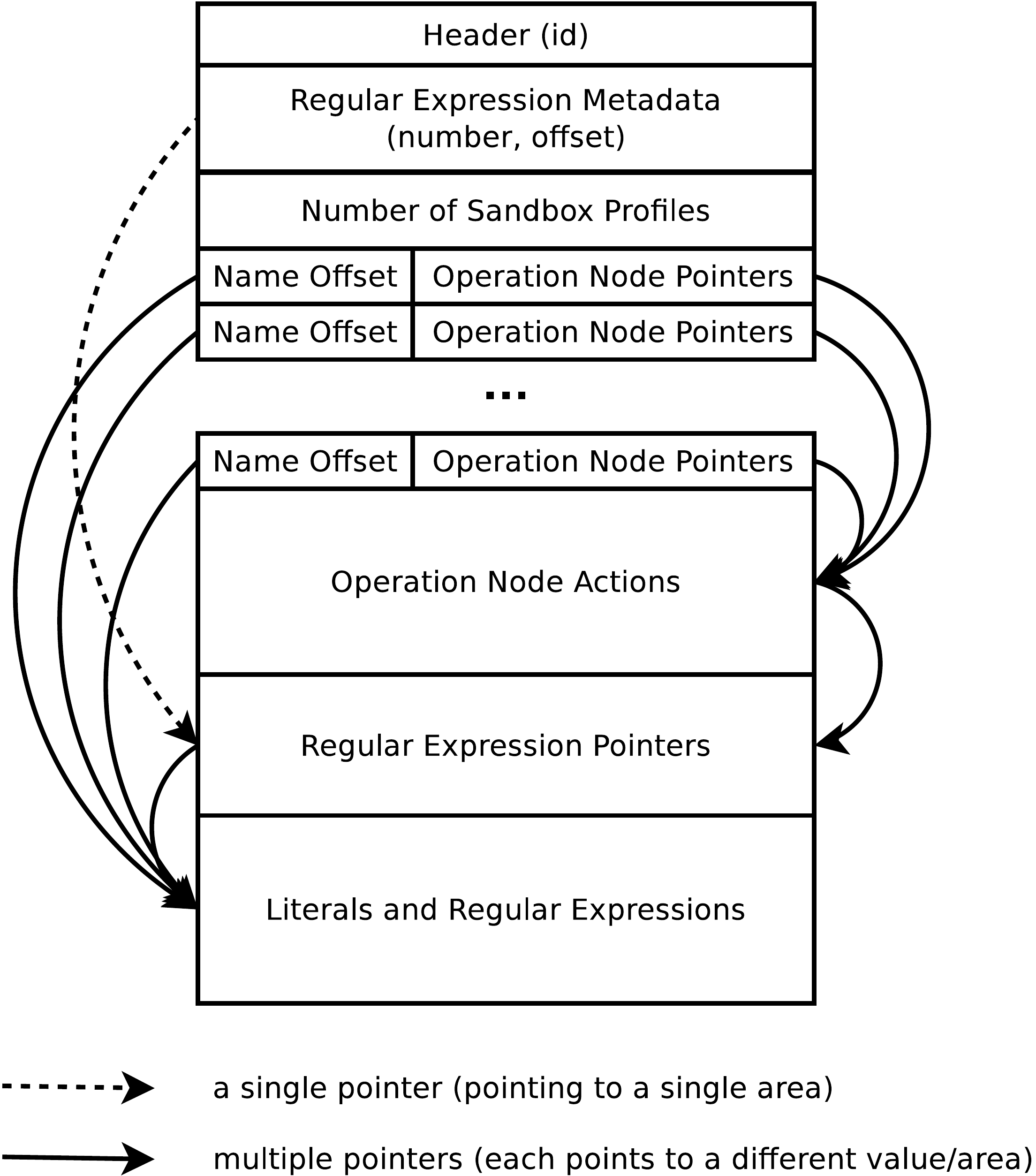}
  \end{center}
  \caption{Overview of the Binary Format of a Sandbox Profile Bundle (iOS 9)}
  \label{fig:binary-format-bundle}
\end{figure}

For the iOS~9 bundled format, sections are aggregated together for all sandbox profiles, as shown in Figure~\ref{fig:binary-format-bundle}. The arrows have the same significance as in Figure~\ref{fig:binary-format}. The difference between the bundled format and the basic separated format are:
\begin{enumerate}
  \item The header id is different. It is \texttt{0x0000} for separated format and \texttt{0x8000} for the bundled format.
  \item Most sections are bundled together for all profiles: \textit{Operation Node Actions}, \textit{Regular Expression Pointers}, \textit{Literals and Regular Expressions}. Information from all sandbox profiles is tied together in a specific section.
  \item The bundled format stores the number of sandbox profiles.
  \item The bundled format aggregates the profile names and stores an offset to the profile name string. \textit{Operation Node Pointers} are stored next to the profile name offset for each sandbox profile.
\end{enumerate}

Reversing a binary sandbox profile requires reversing entries in the \textit{Operation Node Pointers} section to \textit{operations}, entries in the \textit{Operations Node Actions} section to \textit{filters} and \textit{decisions} and entries in the last section of the binary profile (\textit{Literals and Regular Expressions}) to \textit{regular expressions}. We detail these steps in Section~\ref{sec:internals}.

\subsection{The Intermediary Format of Sandbox Profiles}
\label{subsec:intermediary-format}

The sandbox engine transforms the initial SBPL sandbox profile into a Scheme language intermediary format that is then fed to a Scheme interpreter and transformed into the final binary format. Knowledge of the intermediary format is beneficial because of its similarity to the final binary format. We used it when investigating through how \textit{require-any}, \textit{require-all} and \textit{require-not} metafilters are interpreted, as discussed in Section~\ref{sec:internals}. The intermediary format was well documented by Dionysus Blazakis~\cite{Blazakis11theapple} who also built the \textit{apple-scheme} tool~\footnote{\url{https://github.com/dionthegod/XNUSandbox/tree/master/apple-scheme}} for dumping it.

Listing~\ref{lst:intermediary-sandbox-format} shows the intermediary format for the sample sandbox profile in Listing~\ref{lst:simplified-sandbox-profile}. In the intermediary format each operation is assigned a Scheme rule, i.e. a line in Listing~\ref{lst:simplified-sandbox-profile}. The \texttt{default} operation is the first one (index 0) and is always \textit{denied} as shown on line 2. The other operations fallback to the \texttt{default} operation or other operations. The operation with index 7 (on line 9) is \texttt{file*} and the operation with index 14 (on line 16) is \texttt{file-read*}. The \texttt{file-read*} operation has the most complex rule (see line 16) as it is defined explicitly in the sandbox profile in Listing~\ref{lst:simplified-sandbox-profile}; its intermediary rule is a translation of the SBPL rules and, in case none are matched, falls back to the index 7 rule (\texttt{file*}) as shown by the \texttt{(\#f . 7)} construct.

\begin{lstlisting}[caption={Sample Intermediary Sandbox Format},label={lst:intermediary-sandbox-format}]
(
  ((#t deny))
  ((#f . 0))
  ((#f . 0))
  ((#f . 0))
  ((#f . 3))
  ((#f . 3))
  ((#f . 0))
  ((#f . 0))
  ((#f . 7))
  ((#f . 7))
  ((#f . 7))
  ((#t allow))
  ((#f . 7))
  ((#f . 7))
  (((filter path 1 path regex /bin/*) allow) ((filter path #f path literal /bin/secret.txt) deny) (#f . 7))
  ...
)
\end{lstlisting}

\section{Related Work}
\label{sec:related-work}
As a fundamental component in the Apple security layer, the Apple sandbox has been the target of iOS reverse engineers.

In the \textit{iOS Hacker's Handbook}~\cite{miller2012ios}, Miller~et~al. allocated Chapter 5 for sandboxing. Authors of the \textit{iOS Hacker's Handbook} have investigated the Apple sandbox in the past. Dionysus Blazakis was first to describe the internals of the Apple sandbox~\cite{Blazakis11theapple}, and released a set of tools to aid in its analysis. The \textit{container} sandboxing profile for iOS~4 was studied by Dino Dai Zovi~\cite{zovi-blackhat-iOS4}. Information on the language used to construct sandbox profiles was presented in a public white paper~\cite{fgSandboxGuide}. The closest to our work is the reversing work done by Stefan Esser~\cite{esserSandbox}.

Dionysus Blazakis~\cite{Blazakis11theapple} created tools to extract sandbox profiles from iOS~4 and reverse them to a human readable format. However, there are two shortcomings to his work:
\begin{enumerate}
  \item The output is a human readable representation of the inner data structures (i.e., arrays, lists) but not SBPL, i.e. the format of the initial Apple sandbox profile. Not all filters are reversed from their binary values to SBPL tokens. Therefore, we don't have a complete reverser.
  \item Starting with iOS~7, Apple changed the binary format of compiled sandbox profiles, making Blazakis' tools incompatible with recent iOS versions.
\end{enumerate}

Stefan Esser~\cite{esserSandbox} continued Blazakis' work on reversing the Apple Sandbox for more recent iOS versions. Esser solved the second shortcoming, by reversing the newer binary sandbox profile format used in iOS~7 and later. However, the first shortcoming is still present, as Esser released a set of open source tools to extract and decompile built-in Apple iOS binary sandbox profiles into a graph representation (\texttt{.dot} format), not SBPL. Esser also developed a tool for compiling sandbox profiles from SBPL to binary format; we use this to map SBPL tokens to their binary components, as shown in Subsection~\ref{subsec:compile-custom}.

The non-SBPL format output by Essers's and Blazakis' tools has drawbacks for the interested researcher / security professional:
\begin{itemize}
  \item It is not possible to check the tool for correctness. As the output is not SBPL we can not compile it, and we cannot check it for correctness.
  \item The output format is difficult to manually analyze. While it is common to use automated tools to parse and analyze the output format, human inspection and auditing is still expected. Manual inspection of the resulting output for Esser's tools is prohibitive: the generated graphs are large, even for a subset of the sandbox profile. For example, the \texttt{file-read-data} operation from the \textit{container} profile for iOS~7.1.2 is reversed to a graph with 507 nodes and 1009 edges.
  \item The intermediary data structures don't provide support for \textit{metafilters} in the sandbox profiles: \texttt{require-not}, \texttt{require-any}, \texttt{require-all}, further making it difficult to export sandbox profiles in SBPL format.~\footnote{We discuss \textit{metafilters} in Section~\ref{subsec:metafilters}.}
\end{itemize}

We expand on the current state of the art provided by Esser's set of tools and provide SandBlaster as a complete sandbox profile reverser for iOS~7, 8 and 9. SandBlaster performs the non-trivial task of converting the built-in binary representation of sandbox profiles into syntactically correct SBPL. We do this by implementing support for \textit{require-not}, \textit{require-all} and \textit{require-any} metafilters; these are crucial steps in converting a graph representation of binary sandbox profiles (similar to the graph representation provided by Esser) to a human readable SBPL format.

\section{Methodology for Reversing the Apple Sandbox}
\label{sec:methodology}
In this section we show the methodology we employed to extract and reverse Apple iOS binary sandbox profiles. SandBlaster uses this methodology to extract and reverse built-in binary sandbox profiles for iOS~7, 8 and 9. We provide insights and more technical details in Section~\ref{sec:internals}. The methodology can be used to adapt SandBlaster to reverse sandbox profiles for older and newer iOS versions than those already presented.

\begin{figure}[h]
  \begin{center}
    \includegraphics[width=0.5\columnwidth]{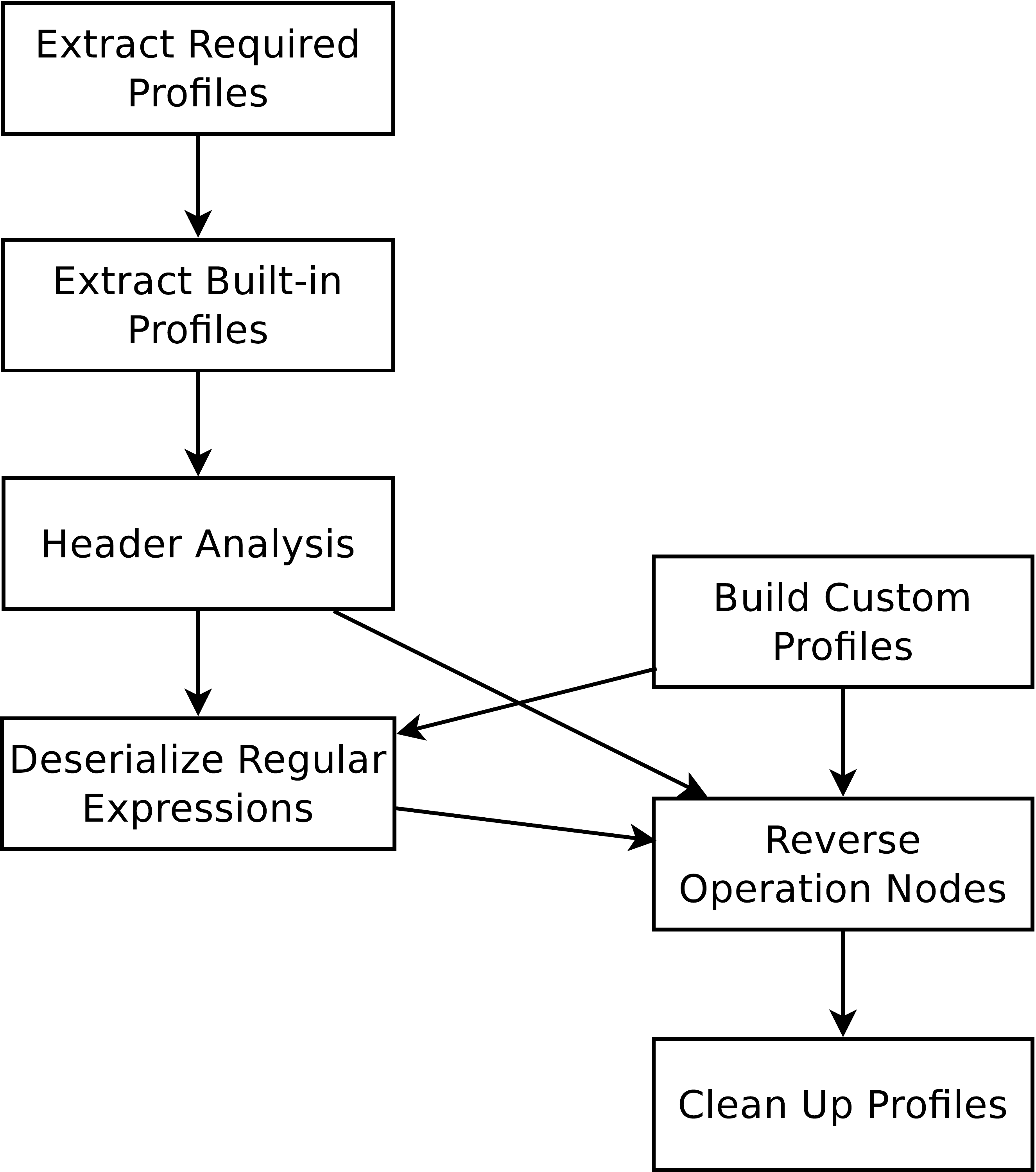}
  \end{center}
  \caption{Methodology for Reversing Apple Sandbox Profiles}
  \label{fig:reverse-methodology}
\end{figure}

Figure~\ref{fig:reverse-methodology} is a high level view of the steps undertaken for reversing binary sandbox profiles. In short, the steps required are:
\begin{itemize}
  \item \textit{Extract Required Files}: In the beginning we extract all required files: files storing the binary profiles and the sandbox operations.
  \item \textit{Extract Built-in Profiles}: Extract the sandbox profile binary blobs.
  \item \textit{Header Analysis}: Extract information about the layout of the binary file storing the binary sandbox profile.
  \item \textit{Build Custom Profiles}: We compile SBPL profiles to their binary format. We use this to create a mapping of plain text sandbox profile tokens to compiled binary components.
  \item \textit{Deserialize Regular Expressions}: Reverse regular expressions from their serialized format to the human readable format. We detail this in Section~\ref{sec:internals}.
  \item \textit{Reverse Operation Nodes}: Map serialized \textit{operations}, \textit{filters}, \textit{metafilters} and \textit{decisions} to their SBPL counterpart. We detail this in Section~\ref{sec:internals}.
  \item \textit{Clean Up Profiles}: Polish the reversed profile to make it as similar to the initial SBPL one as possible. We detail this in Section~\ref{sec:internals}.
\end{itemize}

In the sections below, we do an overview of the first four steps above, leaving the more internally-focused last three steps in Section~\ref{sec:internals}.

\subsection{Extracting Required Files}
\label{subsec:extract-files}

As mentioned in Subsection~\ref{subsec:storing-sandbox-profiles}, between iOS~5 and iOS~8, binary blobs for sandbox profiles were stored in the \texttt{/usr/libexec/sandboxd} file; between iOS~2 and iOS~4 and starting again with iOS~9, sandbox profiles are stored in the sandbox kernel extension (\texttt{com.apple.security.sandbox}). We need to have access to these files to extract the sandbox profile binary blobs. For support in investigating custom sandbox profiles, it is recommended to also have access to the sandbox library file: \texttt{libsandbox.dylib}; the \texttt{libsandbox.dylib} file is part of the \textit{shared library cache}, a bundle of all iOS system libraries, located in \texttt{/System/Library/Caches/com.apple.dyld/dyld_shared_cache_armv7}.

In order to gain access to these files we use publicly available firmware files provided by Apple~\footnote{The XML file available at \url{http://ax.phobos.apple.com.edgesuite.net/WebObjects/MZStore.woa/wa/com.apple.jingle.appserver.client.MZITunesClientCheck/version/} lists the URLs of firmware files for all iOS versions and devices}. The firmware files (using \texttt{.ipsw} extension) are ZIP archive bundles consisting of an encrypted root filesystem image, an encrypted kernel image (also called \textit{kernelcache}) and metadata. Luckily, the iOS reverse engineering community periodically extracts and publishes~\cite{firmware-keys} the root filesystem and kernelcache encryption keys; keys are published shortly after a new iOS version release.

Using dedicated reversing tools (\texttt{vfdecrypt}, \texttt{xpwntool}, \texttt{dmg2img}, \texttt{lzssdec}, \texttt{joker}, \texttt{dsc_extractor}) and information publicly available on the Internet~\cite{kernel-reversing}\cite{shared-cache-extraction}, we created scripts that unpack, decrypt and extract the required files: the \texttt{sandboxd} file, the sandbox kernel extension (\texttt{com.apple.security.sandbox}) and the \texttt{libsandbox.dylib} file. These scripts are part of SandBlaster and provide the automated means of extracting all required files before extracting the built-in binary sandbox profile blobs themselves.

\begin{figure}[h]
  \begin{center}
    \includegraphics[width=0.9\columnwidth]{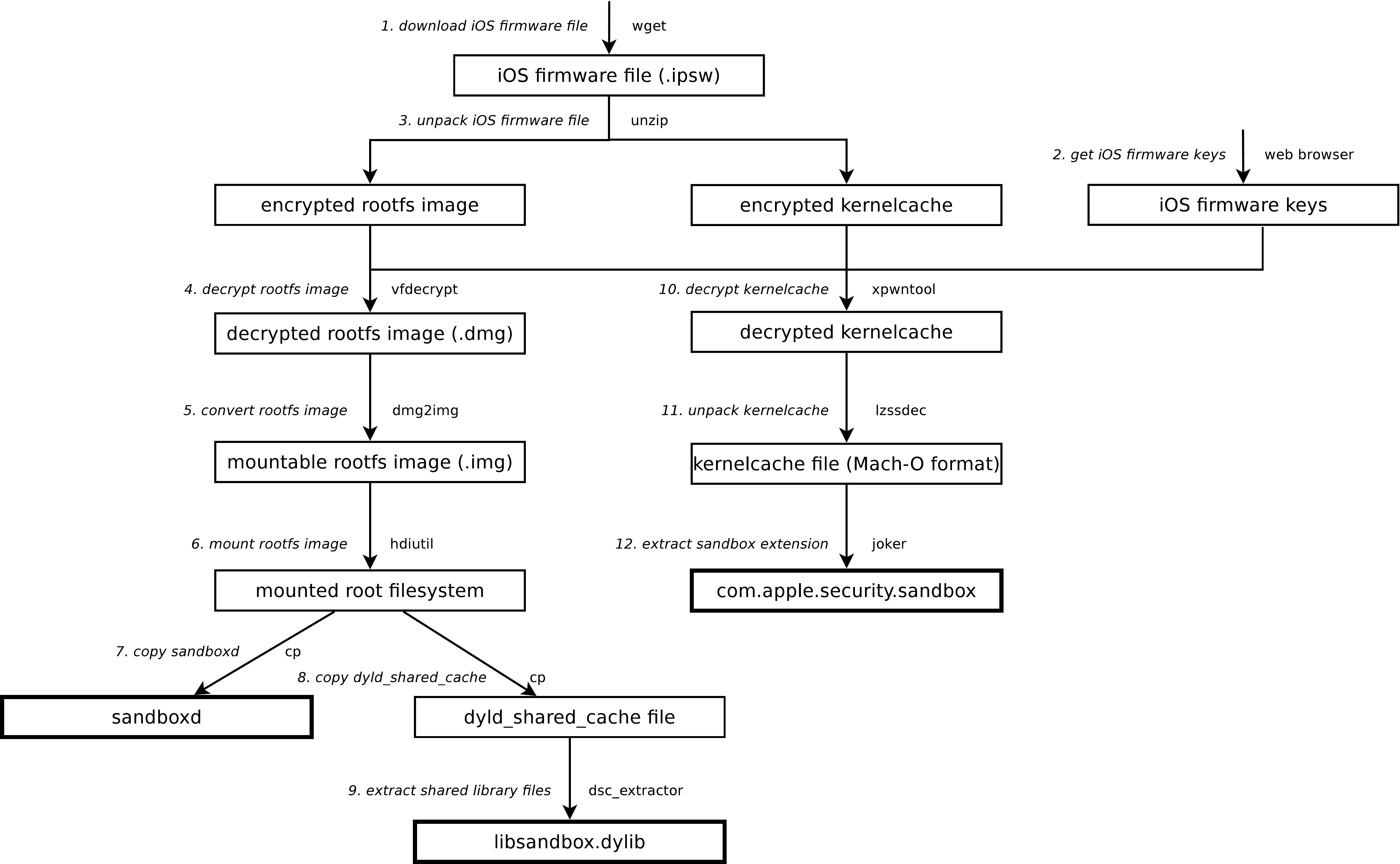}
  \end{center}
  \caption{Process for Extracting Required Files}
  \label{fig:extract-files-process}
\end{figure}

Figure~\ref{fig:extract-files-process} shows the process of extracting required files, the steps where each tool is used and the intermediary results. An arrow corresponds to a step, showing the step description and the tool in use. The following steps are undertaken:

\begin{enumerate}
  \item \textit{download iOS firmware files}: We use the publicly available URLs to download the firmware file for a given iOS version and device. Firmware files use the \texttt{.ipsw} extension and tell the device and iOS version; for example, the firmware file name for iOS 9.3 for iPad2 WiFi is \texttt{iPad2,1_9.3_13E237_Restore.ipsw}.
  \item \textit{get iOS firmware keys}: Reverse engineers manage to extract firmware keys and publish them~\footnote{One such place is ``The iPhone Wiki'': \url{https://www.theiphonewiki.com/wiki/Firmware_Keys}}. We copy the root filesystem and kernelcache keys and map them to the previously downloaded firmware files.
  \item \textit{unpack iOS firmware file}: The firmware file is a ZIP archive. After unpacking it we are provided with the encrypted root filesystem image and the encrypted kernelcache. The encrypted root filesystem image is the largest file with the \texttt{.dmg} extension from the file archive.
  \item \textit{decrypt rootfs image}: We use \texttt{vfdecrypt}~\footnote{\url{https://www.theiphonewiki.com/wiki/VFDecrypt}} and the root filesystem key to decrypt the encrypted root filesystem image. A sample run is shown on line 8 in Listing~\ref{lst:extract-required-files}.
  \item \textit{convert rootfs image}: The decrypted image files consist of multiple partitions. We need to extract the primary partition in IMG format. We use \texttt{dmg2img}~\footnote{\url{http://vu1tur.eu.org/dmg2img}} to locate the primary partition and to convert the root filesystem image (using the \texttt{.dmg} extension) to the mountable root filesystem image (using the \texttt{.img} extension) formatted using HFS+. A sample run is shown on lines 10--14 in Listing~\ref{lst:extract-required-files}.
  \item \textit{mount rootfs image}: We use \texttt{hdiutil}~\footnote{\url{http://commandlinemac.blogspot.ro/2008/12/using-hdiutil.html}} to mount the root filesystem image to a mount point in \texttt{/Volumes} on a Mac OS X filesystem. We could use \texttt{mount} on Linux, but some files are not accessible. A sample run is shown on line 17 in Listing~\ref{lst:extract-required-files}.
  \item \textit{copy sandboxd}: From the mounted volume we copy the \texttt{sandboxd} file.
  \item \textit{copy dyld_shared_cache}: From the mounted volume we copy the \texttt{dyld_shared_cache_file}.
  \item \textit{extract shared library files}: We use \texttt{dsc_extractor}~\cite{shared-cache-extraction} on Mac OS X to extract the library files in the shared cache, particularly the \texttt{libsandbox.dylib} file. A sample run is shown on line 20 in Listing~\ref{lst:extract-required-files}.
  \item \textit{decrypt kernelcache}: We use \texttt{xpwntool}~\footnote{\url{https://github.com/planetbeing/xpwn/tree/master/xpwn}} and the kernelcache key and initialization vector to decrypt the encrypted kernelcache. A sample run is shown on line 32 in Listing~\ref{lst:extract-required-files}.
  \item \textit{unpack kernelcache}: We use \texttt{lzssdec}~\footnote{\url{http://nah6.com/~itsme/cvs-xdadevtools/iphone/tools/lzssdec.cpp}} to unpack the actual kernelcache file (Mach-O format) from the decrypted kernelcache. A sample run is shown on line 35 in Listing~\ref{lst:extract-required-files}.
  \item \textit{extract sandbox extension}: We use \texttt{joker}~\footnote{\url{http://newosxbook.com/tools/joker.html}} to extract the sandbox extension from the kernelcache. A sample run is shown in line 38 in Listing~\ref{lst:extract-required-files}.
\end{enumerate}

All the above steps are automated in SandBlaster. As an overview on how these tools work, Listing~\ref{lst:extract-required-files} shows a sample set of commands one would use manually to extract the three required files (\texttt{sandboxd}, \texttt{libsandbox.dylib} and the sandbox extenstion) from the firmware file \texttt{iPad2,1_9.3_13E237_Restore.ipsw}. The keys for the firmware file are located at \url{https://www.theiphonewiki.com/wiki/Eagle_13E237_\%28iPad2,1\%29}.

\begin{lstlisting}[basicstyle=\fontsize{9}{11}\ttfamily,caption={Sample Commands for Extracting Required Files},label={lst:extract-required-files}]
# Download iOS firmware file for iOS 9.3 on iPad2 WiFi from Apple public URL.
wget http://appldnld.apple.com/iOS9.3/031-54798-20160328-92CDEDBA-F29A-11E5-A70A-B60EF816D560/iPad2,1_9.3_13E237_Restore.ipsw

# Unpack iOS firmware file. The largest extracted .dmg file is the encrypted root filesystem image.
unzip iPad2,1_9.3.1_13E237_Restore.ipsw

# Decrypt the root filesystem image.
vfdecrypt -i058-25537-331.dmg -ke574738d488de59312a53ea9e563ece6300dc9f75029491017a803f9ff52f418025f89af -orootfs.decrypted.dmg

# Find out the primary partition containing the actual filesystem.
dmg2img -l rootfs.decrypted.dmg

# Extract the actual root filesystem from the primary partition (index 3) from the .dmg file.
dmg2img -p 3 -i rootfs.decrypted.dmg -o rootfs.img

# (Mac OS X only) Mount the root filesystem.
hdiutil attach rootfs.img

# Copy the sandboxd file from the mount point. This will fail for iOS 9.3 as there is no sandboxd file for iOS >= 9.
cp /Volumes/Eagle13E237.K93OS/usr/libexec/sandboxd .

# Copy the shared library cache.
cp /Volumes/Eagle13E237.K93OS/System/Library/Caches/com.apple.dyld/dyld_shared_cache_armv7 .

# Extract all library files in the extract-libs/ folder.
dsc_extractor dyld_shared_cache_armv7 extract-libs/

# Copy the libsandbox.dylib file.
cp extract-libs/usr/lib/libsandbox.dylib .

# Decrypt the kernelcache image.
xpwntool kernelcache.release.k93 kernelcache.mach.arm -k 9c48852b0091a3f5081181abb02b5ec58513b15876e7996284398a26265efceb -iv 932342bb7a346ed4494e19d35c76931d -decrypt

# Unpack the kernelcache image. Offset is 448: the start of the 0xfeedface header for Mach-O files.
lzssdec -o 448 < kernelcache.decrypted > kernelcache.mach.arm

# (Mac OS X only) Extract the sandbox extension from the kernelcache.
joker -K com.apple.security.sandbox kernelcache.mach.arm
\end{lstlisting}

\subsection{Extracting Built-in Sandbox Profiles}
\label{subsec:extract-built-in}

As shown in Table~\ref{tab:sandbox-profile-store}, depending on the iOS version, sandbox profiles are stored in binary format either in the \texttt{/usr/libexec/sandboxd} file or in the sandbox kernel extension (\texttt{com.apple.security.sandbox}), depending on the iOS version. Furthermore, each sandbox profile is either located in a distinct location in the file or bundled together with all the other sandbox profiles.

Depending on the storage type, we extract either each sandbox profile one at a time or the sandbox profile bundle. SandBlaster makes use of Stefan Esser's sandbox profile extraction tool (\texttt{extract_sbprofiles}~\footnote{\url{https://github.com/sektioneins/sandbox_toolkit/tree/master/extract_sbprofiles}}) for the separated storage type (iOS $<$ 9) and a custom script to extract the profile bundle for iOS~9. For the custom script for iOS~9, we identify the start of the bundle based on its header and dump data from that point to the end of the sandbox kernel extension file.

After extracting the sandbox profiles, we reverse them to their original SBPL formats. We detail the underlying steps in Section~\ref{sec:internals}.

\subsection{Compiling Custom Sandbox Profiles}
\label{subsec:compile-custom}

In order to reverse filters and regular expressions from the binary sandbox profiles, we need a mapping between their binary representation and their SBPL token. For that, we create custom sandbox profiles in SBPL format and compile them to their binary format. We extract the filters and regular expressions from the binary sandbox profile and we map them to the SBPL tokens in the custom sandbox profiles.

The task of creating custom sandbox profiles faces the challenge of coverage: making sure we cover as much from the SBPL format as possible. This is not trivial as the plain text format for the sandbox profiles is not documented. We rely on two sources of information for this:
\begin{enumerate}
  \item the \texttt{.sb} files located throughout the filesystem on a Mac OS X device, revealing many of the sandbox rules
  \item the \texttt{libsandbox.dylib} library file that stores, among others, plain text descriptions of operations and filters that may be used as rules
\end{enumerate}

The latter has proven to be an invaluable source of information as we were able to determine all operations and all filters that could be used when creating a sandbox profile.

We do extensive testing by creating custom sandbox profiles and then compiling them into the binary format; we then use this to map plain text tokens (\textit{operations}, \textit{filters} or \textit{regular expressions}) to their binary counterparts.

With all operation and filter strings in hand we were able to construct an extensive sandbox profile with all valid operation-filter matches in SBPL expressions. We compiled the plain text format with those rules and obtained the binary format. In order to compile a plain-text format to a binary format, we use the \textit{compile_sb} tool provided by Stefan Esser~\footnote{\url{https://github.com/sektioneins/sandbox_toolkit/tree/master/compile_sb}} that calls the \texttt{sandbox\_compile} library function. We are then able to map plain text expressions (\textit{operations}, \textit{filters} or \textit{regular expressions}) to their binary counterparts; using reverse mapping, we extract plain-text rules from the binary sandbox format.

When compiling custom sandbox profiles we also dump the intermediary format described in Subsection~\ref{subsec:intermediary-format}. We use it to get insight into how the \textit{require-all}, \textit{require-any} and \textit{require-not} metafilters are implemented.

\section{Reversing Internals}
\label{sec:internals}
In this section we show how SandBlaster reverses Apple binary sandbox profiles; we present the steps required for getting a profile in the original SBPL format. We detail the last three steps in Figure~\ref{fig:reverse-methodology}: reversing operation nodes, deserializing regular expressions and cleaning up reversed profiles. We show how we reverse basic operation nodes; then how we reverse the \texttt{require-all}, \texttt{require-any} and \texttt{require-not} metafilters, essential for providing a syntactically correct sandbox profile; we present the steps to deserialize regular expressions; and, finally, what clean up actions are required to get the reversed profile as close to the original SBPL format as possible.

We showed in Section~\ref{sec:sandbox-profiles} that each sandbox profile consists of policy rules applied to operations. Operations define actions performed by an app that are to be checked by the kernel: the kernel maps operations to policy rules defined inside the corresponding binary sandbox profile (for third party apps this would be the \textit{container} profile). Each available operation has a correspondence in the binary format of the sandbox profile. Even if an operation had not been present in the original SBPL format before being compiled, it will still have a correspondence in the binary format, most often linking to the \textit{default} operation decision; e.g., if the \texttt{file-read*} had not been present in the original SBPL format, it will have entries in the \texttt{Operation Node Pointers} and \texttt{Operation Node Actions} sections (as defined in Figure~\ref{fig:binary-format}) and those entries will point to the \textit{default} operation decision (either \textit{allow} or \textit{deny}).

In Section~\ref{sec:sandbox-profiles} we have shown that entries in a sandbox profile consist of operations, filters and decisions. An operation is identified by the operation name, a string. A filter is a key-value pair: the key is a string (or multiple strings), while the value may be a number, a string or another construct such as an IP address. For starters, we need to have a list of all operations and of all filter keys, with their corresponding binary representation, and all possible mappings between operations and filters. We extracted this information from the \texttt{libsandbox.dylib} file and then constructed human readable to binary counterpart mappings as shown in Subsection~\ref{subsec:compile-custom}.

\subsection{Reversing Operation Nodes}
\label{subsec:reverse-operation-nodes}

For each operation, the sandbox profile defines a set of rules. A rule may be a decision (\textit{allow} or \textit{deny}) or a filter together with a decision. Rules may be affected by metafilters: \textit{require-any}, \textit{require-all} or \textit{require-not}. A rule is serialized in the binary profile in an entry that we call an \texttt{operation node} or a \texttt{node}. For each operation there is an array of \textit{operation nodes} in the \textit{Operation Node Actions} section in Figure~\ref{fig:binary-format}. The \textit{Operation Node Actions} section concatenates the arrays of operation nodes for each operation; the first item in the operation node array for each operation is pointed to by an entry in the \textit{Operation Node Pointers} section.

There are two types of nodes: terminal nodes and non-terminal nodes. We will discuss them shortly. Most of the nodes are non-terminal nodes. Each non-terminal node consists of:
\begin{itemize}
  \item \textbf{serialized filter key}: the binary counterpart of the filter key;
  \item \textbf{serialized filter value}: the binary counterpart of the filter value;
  \item \textbf{match node offset}: the offset to the next node to be checked in case the filter is matched;
  \item \textbf{unmatch node offset}: the offset to the next node to be checked in case of the filter is not matched.
\end{itemize}

As each non-terminal node defines a match node offset and an unmatch node offset (links to other nodes), we conclude that each operation is defined as a series of nodes joined together in an directed acyclic graph. We now define terminal nodes and non-terminal nodes:
\begin{itemize}
  \item \textbf{terminal nodes} are end nodes (leaves) in the operation node graph (i.e. nodes that are not the starting point of any edge)
  \item \textbf{non-terminal nodes} are intermediary nodes in the operation node graph (i.e. nodes that are the starting point for edges)
\end{itemize}

We also define two types of edges:
\begin{itemize}
  \item \textbf{match edge} is an edge that is defined by a \textbf{match node offset} in the operation node
  \item \textbf{unmatch edge} is an edge that is defined by a \textbf{unmatch node offset} in the operation node
\end{itemize}

\begin{figure}[h]
  \begin{center}
    \includegraphics[width=0.7\columnwidth]{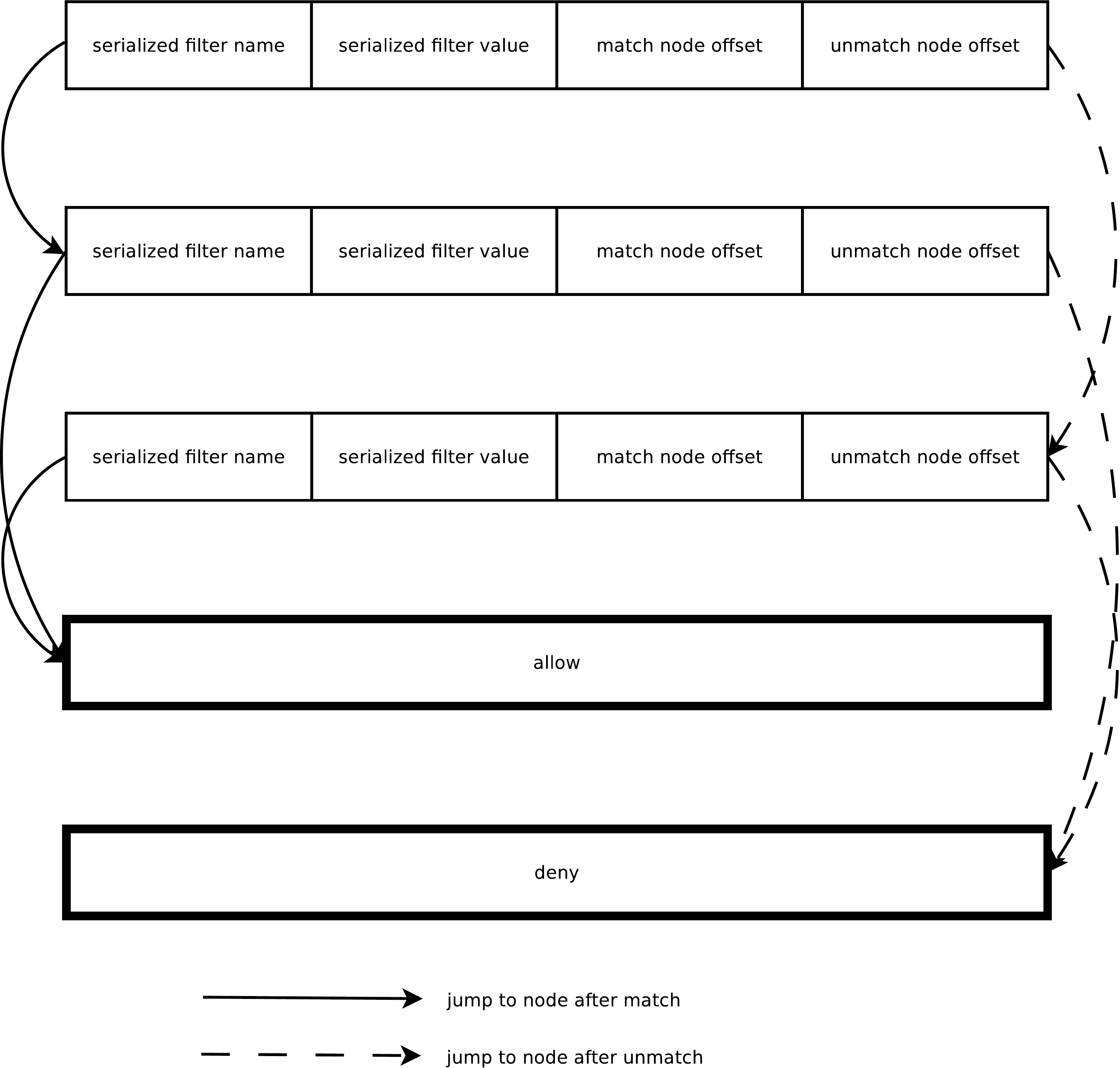}
  \end{center}
  \caption{Graph-like View for Operation Node Array}
  \label{fig:operation-node-array}
\end{figure}

Figures~\ref{fig:operation-node-array},~\ref{fig:require-not-graph},~\ref{fig:node-removal} use the following convention for representing the above node and edge types:
\begin{itemize}
  \item Non-terminal nodes are marked as simple border components.
  \item Terminal nodes use a thicker border. These are nodes named \textit{allow} and \textit{deny}.
  \item Match edges are represented as normal arrows.
  \item Unmatch edges are represented as dashed arrows.
\end{itemize}

Each operation uses the corresponding pointer in the \textit{Operation Node Pointers} and identifies the starting node in its array of operation nodes; it then starts matching each node. In Figure~\ref{fig:operation-node-array} we show how nodes are bound together in an directed acyclic graph for a given operation. If a match is found it follows a match edge to the node pointed to by the \texttt{match offset}; otherwise it follows an unmatch edge to the node pointed to by the \texttt{unmatch offset}. At some point it will reach one of the two terminal nodes: the \textit{allow} terminal node or the \textit{deny} terminal node. The \textit{allow} and \textit{deny} terminal nodes are always present in a binary sandbox profile: if the \textit{allow} node is reached in the graph, the operation succeeds, if the \textit{deny} node is reached in the graph, the operation fails.

As the sandbox profile uses a default decision, we need only consider the edges corresponding to the opposite decision; i.e. in case of the \textit{deny} default decision, we need only consider edges leading to the \textit{allow} node; edges leading to the \textit{deny} node need not be considered, as they they fall back to the \textit{deny} default decision. So our goal is to only consider paths in the direct acyclic graph that lead to the \textit{allow} terminal node in case of a \textit{deny} default decision.

\subsubsection{Reversing a Simple Operation Node Array}
\label{subsubsec:reverse-simple-operation-node}

Assuming we compiled the sandbox profile from Listing~\ref{lst:sample-profile} we extract the operation node array corresponding to the \texttt{file-read*} operation. The array is shown in hexadecimal in Listing~\ref{lst:hex-operation-node-array} and each item broken up in Table~\ref{tab:operation-node-fields}~\footnote{The format is little endian, such that a \texttt{2300} representation corresponds to the number \texttt{0x0023}.} We extract the array by following the offset corresponding to the \texttt{file-read*} operation in the \textit{Operation Node Pointers} section.

\begin{lstlisting}[caption={Sample Profile},label={lst:sample-profile}]
(version 1)
(deny default)
(allow file-read*
    (regex #"/bin/*")
    (vnode-type REGULAR-FILE))
\end{lstlisting}

\begin{lstlisting}[caption={Operation Node Array in Hexadecimal},label={lst:hex-operation-node-array}]
0081 0000 2300 2200
001d 0100 2300 2400
\end{lstlisting}

\begin{table}[h]
  \begin{center}
  \begin{tabular}{lllll}
    \toprule
    \textbf{Node Type} & \textbf{Filter Key} & \textbf{Filter Value} & \textbf{Match Offset} & \textbf{Unmatch Offset} \\
    \midrule
    0x00 & 0x81 & 0x0000 & 0x0023 & 0x0022 \\
    0x00 & 0x1d & 0x0001 & 0x0023 & 0x0024 \\
    \bottomrule
  \end{tabular}
  \end{center}
  \caption{Break-Up of Operation Node Fields}
  \label{tab:operation-node-fields}
\end{table}

We do a mapping of filter keys and filter values to the SBPL equivalent; and we follow the pointers from the match offset and unmatch offset fields to other operation nodes. We do the following conversions:
\begin{itemize}
  \item The node type for both operation nodes is \texttt{0x00} meaning this is a non-terminal node. What follows is then a filter definition.
  \item The first serialized filter key is \texttt{0x81}. This corresponds to the \textit{regex} filter key.
  \item The binary value for the second serialized filter key is \texttt{0x1d}. This corresponds to the \textit{vnode-type} filter key.
  \item The first serialized filter value is \texttt{0x0000}. This is the index of the regular expression; we are using the first regular expression in the binary sandbox profile (index \texttt{0}). We reverse the first serialized regular expression to \textit{/bin/*}.
  \item The second serialized filter value is \texttt{0x0001}. This corresponds to the \textit{REGULAR-FILE} filter value.
  \item For both nodes, if the filters are matched, the next node is referenced by the \texttt{0x0023} offset. This value is multiplied by 8 for the actual offset (\texttt{0x118}) pointing to the next node. The node at offset \texttt{0x118} is the \textit{allow} terminal node. So for both nodes, if filters are matched, the decision is \textit{allow}.
  \item For the first node, if the filter is \textbf{not} matched, the next node is referenced by the \texttt{0x0022} offset. That value corresponds to the second node; so we have a fallback: in case the first node is not matched, we fall back to checking the second node.
  \item For the second node, if the filter is \textbf{not} matched, the next node is reference by the \texttt{0x0024} offset. This value is multiplied by 8 for the actual offset (\texttt{0x120}) pointing to the next node. The node at offset \texttt{0x120} is the \textit{deny} terminal node. So in case the filter for the second node is not matched, the decision is \textit{deny}. Assuming the fallback, in case neither of the two nodes is matched, the decision is \textit{deny}.
\end{itemize}

Based on the above conversions, Listing~\ref{lst:operation-node-array-pseudocode} shows the equivalent pseudocode for the two rules in the operation node array for the \texttt{file-read*} operation. This pseudocode corresponds to a graph representation of the rules. Based on the graph representation we derive the original SBPL format represented in Listing~\ref{lst:sample-profile}.

\begin{lstlisting}[caption={Operation Node Array in Hexadecimal},label={lst:operation-node-array-pseudocode}]
if (match_regex("/bin/*"))
    return allow;
elif (match_vnode_type(REGULAR-FILE))
    return allow;
return deny;
\end{lstlisting}

\subsubsection{Reversing the require-not Metafilter}
\label{subsubsec:reverse-require-not}

The simplest non-terminal node would have the following form: if it matches, do a \texttt{deny}/\texttt{allow}, if it doesn't match, do the opposite. There are however intermediary non-terminal nodes. The existence of intermediary non-terminal nodes is caused by the presence of rule modifiers or metafilters: \textit{require-not}, \textit{require-all} and \textit{require-any}. These rule modifiers are the equivalent of a logical not, logical and logical or, respectively. To make matters more complicated, they may be nested.

The format of non-terminal rules and the rule for the \textit{default} operation dictate the use of the \textit{require-not} metafilter. For any rule, the \textit{match node} (i.e. the next node from the \textit{match edge}) and \textit{unmatch node} (i.e. the next node from the \textit{unmatch edge}) may fit into one of the cases featured in Table~\ref{tab:match-unmatch-rules}. Assuming the \textit{default} operation uses the \textit{deny} decision, we want to get to a situation where \textit{match nodes} may only be the \textit{allow} terminal node or a non-terminal node, while \textit{unmatch nodes} may only be a \textit{deny} terminal node or a non-terminal node. This will allow us to only consider \textit{match edges} in the graph. If that is not the case we negate the node and reverse the \textit{match node} with the \textit{unmatch node}; a node is negated using the \textit{require-not} metafilter. In the end, we have a \textit{match graph} for each operation, i.e. a graph where we only need to follow \textit{match edges} to get to the \textit{allow} terminal node (again, assuming the \textit{default} operation uses a \textit{deny} decision).

\begin{table}[h]
  \begin{center}
  \begin{tabular}{lll}
    \toprule
    \textbf{Match Node} & \textbf{Unmatch Node} & \textbf{Action} \\
    \midrule
    allow & deny & keep the same \\
    deny & allow & \textbf{negate} \\
    allow & non-terminal & keep the same \\
    deny & non-terminal & \textbf{negate} \\
    non-terminal & allow & \textbf{negate} \\
    non-terminal & deny & keep the same \\
    non-terminal & non-terminal & keep the same \\
    \bottomrule
  \end{tabular}
  \end{center}
  \caption{Negate Action for Match/Unmatch Nodes (\textit{default} operation uses \textit{deny} decision)}
  \label{tab:match-unmatch-rules}
\end{table}

This approach takes care of the \textit{require-not} metafilter. Each time we use a \textbf{negate} action as show in Table~\ref{tab:match-unmatch-rules} we prefix the current rule with a \textit{require-not (\ldots)} construct.

\begin{figure}[h]
  \begin{center}
    \includegraphics[width=0.7\columnwidth]{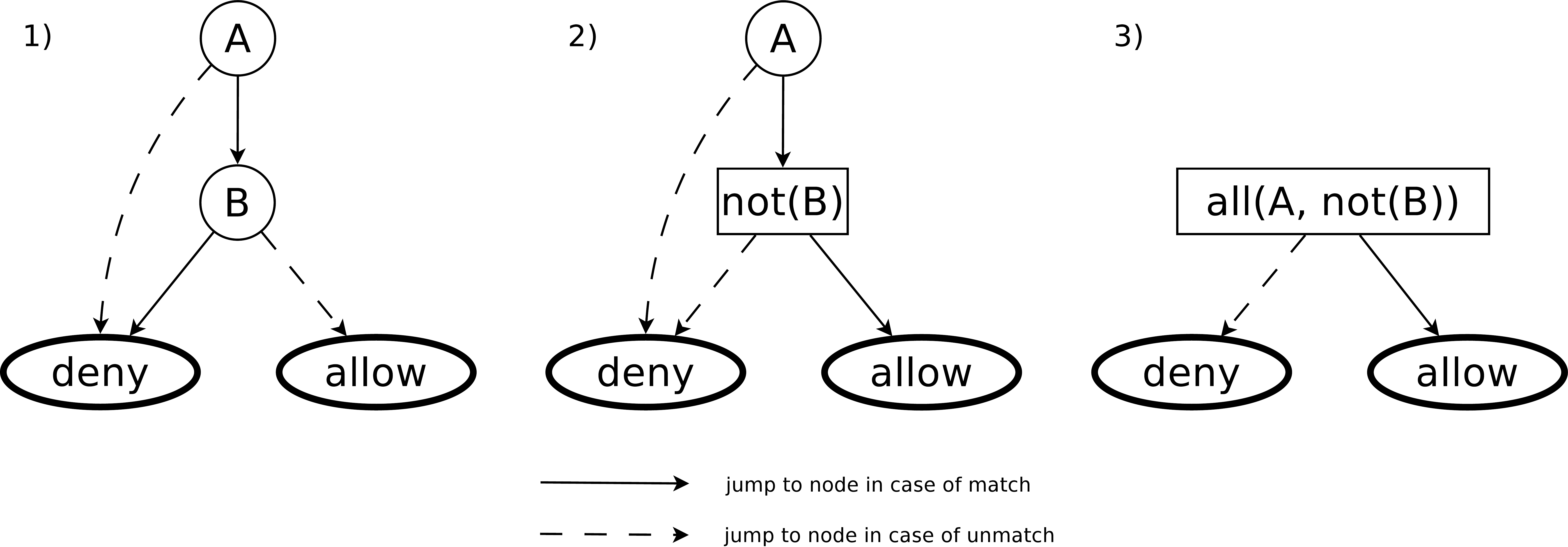}
  \end{center}
  \caption{Reducing Graph with require-not Metafilter}
  \label{fig:require-not-graph}
\end{figure}

Figure~\ref{fig:require-not-graph} shows a simple situation where the \textit{require-not} and \textit{require-all} metafilters are used. Operation node B falls in the second situation from Table~\ref{tab:match-unmatch-rules}, where a match results in a \textit{deny} decision. In the second step in Figure~\ref{fig:require-not-graph} we negate node B by applying the \textit{require-not} metafilter (dubbed \textit{not(B)} in the figure for brevity); after negating the node, it now points to the \textit{allow} node in case of a match. In the third step we use the \textit{require-all} metafilter (dubbed \textit{all(\ldots)} in the figure for brevity) to state that \textbf{both} node A and \texttt{require-not(B)} need to be matched for an \textit{allow decision}.

\subsubsection{Reversing Nested Rules}
\label{subsubsec:reverse-nested-rules}

\textit{require-any} and \textit{require-all} metafilters may aggregate multiple nodes and even nest other \textit{require-any} and \textit{require-all} metafilters. In Listing~\ref{lst:require-all}, the \textit{require-all} metafilter forces the \textit{allow} decision to be taken only if \textbf{both} the \texttt{regex} filter and \texttt{extension} filter are matched.

\begin{lstlisting}[caption={Sample Usage of the require-all Metafilter},label={lst:require-all}]
(allow file-read*
    (require-all
        (regex "^/dev/ttys[0-9]*")
        (extension "com.apple.sandbox.pty")))
\end{lstlisting}

\begin{figure}[h]
  \begin{center}
    \includegraphics[width=0.5\columnwidth]{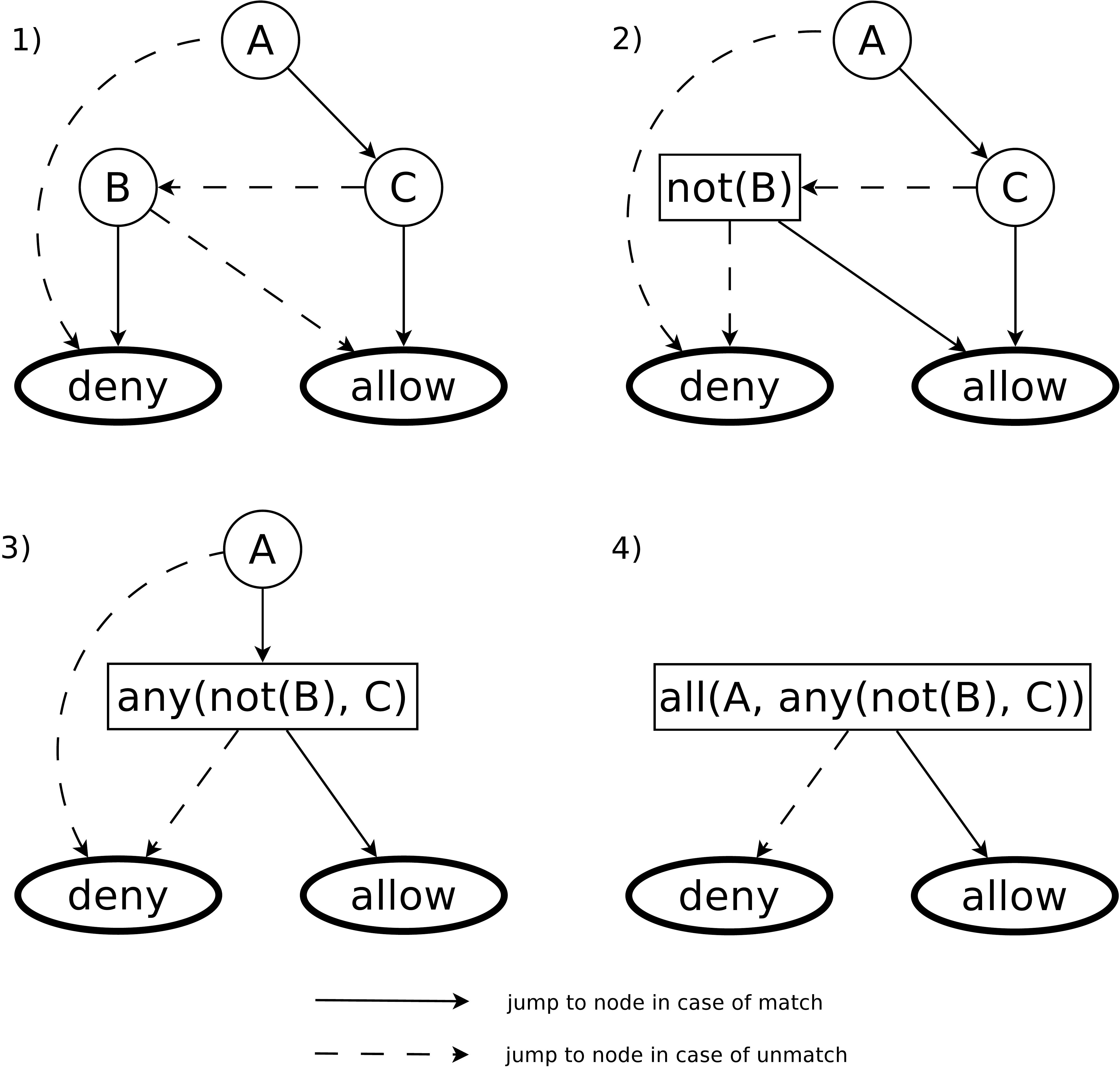}
  \end{center}
  \caption{Reversing Nested Rules through Node Removal}
  \label{fig:node-removal}
\end{figure}

Once we have a directed acyclic graph representation of the rules in the binary sandbox profile (some including the \textit{require-not} metafilter), we make use of node removal (or node aggregation) to construct the \textit{require-all} and \textit{require-any} metafilters. We first aggregate together sibling nodes into \textit{require-any} metafilters and then we aggregate parent and child nodes into \textit{require-all} metafilters. Figure~\ref{fig:node-removal} shows a sample rule graph consisting of three nodes (A, B and C) that has been built from nested \textit{require-not}, \textit{require-any} and \textit{require-all} metafilters. Each step apart from the first one alters the graph through the use of metafilters:
\begin{enumerate}
  \item The first phase is the initial graph. The starting point is node B pointing to the \textit{deny} node in case of match.
  \item In step 2, we reverse the decision of node B by using the \textit{require-not} metafilter (\texttt{not} in the figure) and we now have a new node (\texttt{not(B)}) that points to the \textit{allow} node in case of a match.
  \item In step 3, we see that \textbf{either} node C or node \texttt{not(B)} point to the \textit{allow} node. We use the \textit{require-any} metafilter (\texttt{any} in the figure) to unite the two nodes into a new node: \texttt{any(not(B), C)}.
  \item Finally, in step 4, we need \textbf{both} node A and node \texttt{any(not(B), C)} to reach the \textit{allow} node. We use the \textit{require-all} metafilter (\texttt{all} in the figure) to unite the two nodes into a final node: \texttt{all(A, any(not(B), C))}.
\end{enumerate}
Afterwards, the final nodes are then translated to the original SBPL format.

\subsection{Deserializing Regular Expressions}
\label{subsec:reverse-regex}

Regular expressions are an important part of the sandbox profile; the container profile for iOS 8.4.1 uses 1964 filter nodes out of which 131 (about 6.6\%) use regular expressions. Unlike literals in filters, which are stored as strings inside the binary format, regular expressions are serialized (or binarized) in the binary format. The binary representation of the regular expressions is that of a serialized non-deterministic finite automaton (\textit{NFA}), as pointed out by Stefan Esser~\cite{esserSandbox}.

There are two steps to undertake when reversing the serialized format of the regular expressions:
\begin{enumerate}
  \item build the NFA from the binary format (reverse it);
  \item convert the NFA to the original regular expression.
\end{enumerate}

In order to reverse the binary representation of the NFA to a graph-like representation we need to map between characters and metacharacters in a regular expression to those in the serialized format. As in the case of operation node filters, we created a series of rules with many regular expressions in a sandbox profile file and then compared that to the binary counterpart. We extracted the binary counterpart for meta-characters such as any character, beginning of line, ending of line, character class, and meta-nodes such as jump forward, jump backward. Jump forward and jump backward are intermediary nodes pointing to other nodes; they help construct the $\epsilon$-transitions in the NFA representation of the regular expression: jump backwards connects the current node to a previously encountered node (using a lower index) in the NFA while jump forward connects it to ``future'' node (with a higher index) in the NFA. With this information we traversed the binary representation and constructed a graph representation for the NFA, similar to the way NFAs are usually represented.

\begin{figure}[h]
  \begin{center}
    \includegraphics[width=0.7\columnwidth]{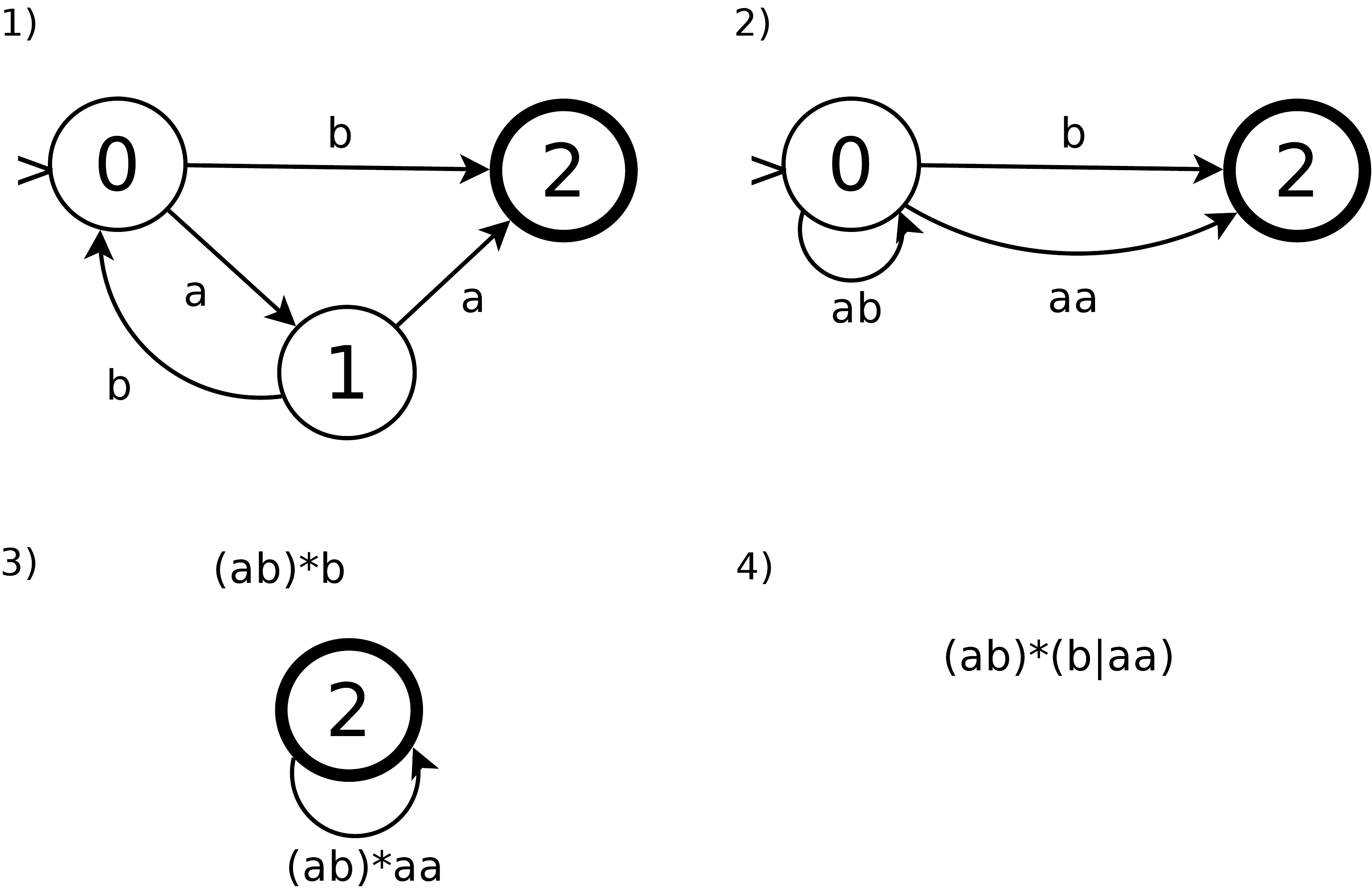}
  \end{center}
  \caption{Sample Steps to Reverse an Automaton into a Regular Expression}
  \label{fig:state-removal}
\end{figure}

With the NFA extracted, we need to transform it into a regular expression. For this, we used a state removal approach: take one node and remove it, making sure that all its previous nodes are now connected to its next nodes in the graph with proper string information on the transitions. Figure~\ref{fig:state-removal} shows a sample automaton and the steps undertaken to retrieve the original regular expression. We use state removal at each step and update transitions between remaining nodes. This is not a classical representation of an NFA as we have regular expression on transitions, whereas a classical representation uses only characters from the alphabet.

With the serialized regular expressions now reversed, they are fed back to the operation node action filters. We now have a complete human readable form of the binary sandbox profile, very close to the initial one. Several clean up steps need to be taken care of to polish the profile, as discussed in the next section.

\subsection{Cleaning Up Reversed Profiles}
\label{subsec:clean-up}

When a sandbox profile is compiled, there is a set of implicit SPBL rules set that are added to it; these rules are serialized in the binary sandbox profile. We need to remove these rules from the reversed sandbox profile to make it as close as possible to the original SBPL one. Listing~\ref{lst:implicit-sbpl-rules} shows the implicit rules that are added to each sandbox profile for iOS 9.3; they are stored in plain text in the \texttt{libsandbox.dylib} file.

\begin{lstlisting}[caption={Implicit SBPL Rules for Sandbox Profiles},label={lst:implicit-sbpl-rules}]
;;;;;
;;;;; Standard policy applied to all sandboxed processes.
;;;;;
;;;;; Copyright (c) 2014 Apple Inc. All rights reserved.
(version 1)
(define (allowed? op)
  (sbpl-operation-can-return? op 'allow))
(define (denied? op)
  (sbpl-operation-can-return? op 'deny))
;; Allow mach-bootstrap if mach-lookup is ever allowed.
(if (allowed? mach-lookup)
  (allow mach-bootstrap))
;; Allow access to webdavfs_agent if file-read* is always allowed.
;; <rdar://problem/6816031> remove workaround for 6769092
(if (not (denied? file-read*))
  (allow network-outbound
         (regex #"^/private/tmp/\.webdavUDS\.[^/]+$")))
;; Never allow a sandboxed process to open a launchd socket.
(deny network-outbound
      (literal "/private/var/tmp/launchd/sock")
      (regex #"^/private/tmp/launchd-[0-9]+\.[^/]+/sock$"))
;; Always allow a process to signal itself.
(allow signal (target self))
\end{lstlisting}

Based on information from \texttt{libsandbox.dylib} and by building custom sandbox profiles, we identify implicit SBPL rules and we follow two steps in cleaning up reversed profiles: removing extra operation rules and removing implicit regular expression rules.

For the first step, we remove operations whose decision is identical to the default one (\textit{deny} or \textit{allow}) and operations that we know have been added implicitly (such as \texttt{(allow signal (target self))}, as seen in line 23 in Listing~\ref{lst:implicit-sbpl-rules}).

Certain regular expression patterns (such as \texttt{\textasciicircum/private/tmp/launchd-$[$0-9$]$+.$[$\textasciicircum/$]$+/sock\$}) are compiled inside the binary format for given operations. We discard these regular expressions constructs from the reversed format.

In the end we are able to reverse built-in Apple binary sandbox profiles into human readable SBPL files. We compiled the resulting sandbox profiles and validated their syntactical correctness with minor to no modifications. As we compiled the resulted profiles on Mac OS X, minor modifications were required to the reversed sandbox profiles, as they were using tokens that differ between iOS and Mac OS X.

\section{Conclusion and Further Work}
\label{sec:conclusion}
We presented SandBlaster, a software bundle that decompiles built-in Apple binary sandbox profiles to their original human readable SBPL (\textit{Sandbox Profile Language}) format. We built our tool based on previous work by Dionysus Blazakis and Stefan Esser and enhanced their work by providing an easy to analyze, fully human readable format. 

We reversed the built-in binary sandbox profiles for iOS 7, 8 and 9. We were able to recompile reversed sandbox profiles with minimal modifications due to inherent differences between iOS and Mac OS X (the compilation platform) and thus prove their syntactic correctness. The reversed profiles are useful for studying the inner workings of iOS sandbox profiles, particularly the default \textit{container} profile used for third party iOS apps.


SandBlaster is a software bundle consisting of original software, existing tools and scripts that automate the extraction and reversing process such that, given an iOS firmware file, we use only few commands to reverse all sandbox profiles for that particular iOS version.

In the future, we aim to validate the semantic correctness by running apps and mapping them to the reversed SBPL sandbox profiles, particularly the \textit{container} profile. This requires a jailbroken iOS device and mapping a given app to a custom sandbox profile.

\bibliographystyle{abbrv}
\bibliography{sandbox-decompiler}

\begin{thebibliography}{10}

\bibitem{shared-cache-extraction}
{ant4g0nist}.
\newblock {iOS Shared Cache Extraction to solve "redacted" problem}.
\newblock
  \url{http://ant4g0nist.blogspot.ro/2015/04/ios-shared-cache-extraction-to-solve.html}.
\newblock Accessed: 2016-07-04.

\bibitem{iOS9White}
Apple.
\newblock {iOS Security Guide}.
\newblock \url{https://www.apple.com/business/docs/iOS_Security_Guide.pdf}.
\newblock Accessed: 2016-04-11.

\bibitem{apple-app-sandbox-design-guide}
{Apple Inc.}
\newblock {App Sandbox Design Guide}.
\newblock
  \url{https://developer.apple.com/library/mac/documentation/Security/Conceptual/AppSandboxDesignGuide/AboutAppSandbox/AboutAppSandbox.html}.
\newblock Accessed: 2016-07-24.

\bibitem{Blazakis11theapple}
D.~Blazakis.
\newblock {The Apple Sandbox}.
\newblock In {\em {In Black Hat DC}}, 2011.

\bibitem{esserSandbox}
S.~Esser.
\newblock {iOS8 Containers, Sandboxes and Entitlements}.
\newblock
  \url{http://www.slideshare.net/i0n1c/ruxcon-2014-stefan-esser-ios8-containers-sandboxes-and-entitlements}.
\newblock Accessed: 2016-04-11.

\bibitem{fgSandboxGuide}
fG!
\newblock {Apple's Sandbox Guide v1.0}.
\newblock
  \url{http://reverse.put.as/wp-content/uploads/2011/09/Apple-Sandbox-Guide-v1.0.pdf}.
\newblock Accessed: 2016-04-11.

\bibitem{ios-popularity}
{IDC Research, Inc.}
\newblock {Smartphone OS Market Share, 2015 Q2}.
\newblock \url{http://www.idc.com/prodserv/smartphone-os-market-share.jsp}.
\newblock Accessed: 2016-07-24.

\bibitem{miller2012ios}
C.~Miller, D.~Blazakis, D.~DaiZovi, S.~Esser, V.~Iozzo, and R.-P. Weinmann.
\newblock {\em {iOS Hacker's Handbook}}.
\newblock John Wiley \& Sons, 2012.

\bibitem{kernel-reversing}
{NowSecure}.
\newblock {iOS Kernel Reversing Step by Step}.
\newblock
  \url{https://www.nowsecure.com/blog/2014/04/14/ios-kernel-reversing-step-by-step/}.
\newblock Accessed: 2016-07-04.

\bibitem{firmware-keys}
{the iPhone wiki}.
\newblock {Firmware Keys}.
\newblock \url{https://www.theiphonewiki.com/wiki/Firmware_Keys}.
\newblock Accessed: 2016-07-04.

\bibitem{zovi-blackhat-iOS4}
D.~D. Zovi.
\newblock {Apple iOS 4 Security Evaluation}.
\newblock In {\em BlackHat USA}, 2011.

\end{thebibliography}

\end{document}